\def\bey{\begin{eqnarray}}
\def\eey{\end{eqnarray}}
\def\be{\begin{equation}}
\def\ee{\end{equation}}
\def\gm{\gamma}

\def\Ld{\Lambda}
\def\af{\alpha}
\def\sg{\sigma}

\def\om{\omega}
\def\r{\rho}

\def\bt{\beta}

\def\pp{\partial}
\def\pp{\partial}

\def\nnb{\nonumber}
\documentstyle[11pt,epsfig]{article}
\title{Charge density of light exotic nuclei and $\r NN$ tensor coupling}
\author{ W. Z. Jiang\footnote{Email: jiangwz02@hotmail.com}, Y.L.Zhao \\
  $^1$ Shanghai Institute of Applied Physics,\\
  Chinese Academy of Sciences,Shanghai 201800,China\\
  $^2$ Center of Theoretical Nuclear Physics,\\
 National  Laboratory of Heavy Ion Accelerator, Lanzhou
 730000,China
 }
\date{}
\bigskip
\begin{document}
\maketitle
\baselineskip 20.6pt
\begin{abstract}
\baselineskip 18.0pt
 We use a relativistic mean field model to study the charge
 density distributions of exotic oxygen isotopes. Nonlinear
 isoscalar-isovector terms which are not constrained by the
 present data are considered and  the  $\r NN$ tensor
 coupling  that does not affect the symmetry energy in the mean
field model is included.  Strong correlations between the neutron
radius of $^{208}$Pb and the charge radius of $^{12,13,23,24}$O
are found. The $\r NN$ tensor coupling explicitly enhances the
radius correlations  between  $^{23,24}$O and $^{208}$Pb. This
enhancement is due to the neutron occupation of the orbital
$2s_{1/2}$. The charge radius is sensitive to the change of the
density dependence of the symmetry energy as the isotope goes
towards the proton drip line.

\end{abstract}



\newpage
The nuclear equation of state (EOS) is  fundamentally important to
describe the nuclear phenomenology. The EOS comprises of two
components: the saturation property of symmetric matter and the
asymmetric energy. There exists a Cooster band for the saturation
property of symmetric matter. However, the density dependence of
the symmetric energy is much more poorly known to
date\cite{brown,li}. Therefore, the understanding of amount of
properties of neutron stars, nuclei far from the $\bt$ stability,
and radioactive ion collisions exists theoretical uncertainties.
Many factors, for instance, the isoscalar-isovector coupling
terms\cite{horo1,horo2,horo3,horo4,todd}, the isovector-scalar
mesons and density dependent coupling constants\cite{ma,alon}, can
modify the behavior of the density dependence of the symmetry
energy. Recently the density dependence of the symmetry energy has
been extensively explored through the inclusion of the
isoscalar-isovector coupling terms.

The inclusion of the isoscalar-isovector coupling terms allows one
modify the neutron skin of heavy nuclei without changing the
saturation property of symmetric matter and  a variety of
ground-state properties that are well constrained
experimentally\cite{horo1}. Theoretical estimates may give an
uncertainty of the neutron radius of $^{208}$Pb by about 0.3fm.
Recently,  data to data correlations between the neutron radius of
$^{208}$Pb and neutron star properties have been
established\cite{horo1,horo2,horo3,horo4}. Such correlations are
also found in the binding energy of valence neutrons for some
neutron rich nuclei\cite{todd}. Accordingly, an accurate
measurement of the neutron radius for $^{208}$Pb at the Jefferson
Laboratory\cite{jeff} that promises a 1\% accuracy is very
significant.

With the development of the radioactive beams properties of many
exotic nuclei have been explored\cite{jens}. Many neutron and
proton halo nuclei are identified through the measurement of the
reaction cross sections and longitudinal momentum distributions of
nucleus breakup. To obtain  more definite information of halo
nuclei, a pure probe that excludes the strong interaction is
necessary. At RIKEN\cite{riken} and GSI\cite{gsi}, new facilities
that provide electron scattering on exotic nuclei are under
construction. These facilities will provide an elaborate
explorations on the charge density distributions of proton halo
nuclei. Eventually, the charge density distributions of all exotic
nuclei are yet to be surely measured.  Recently, Wang et.al.
investigated elastic electron scattering on light exotic
proton-rich nuclei theoretically\cite{wang}. On the other hand,
the theoretical charge density distributions for these nuclei may
exist uncertainties since the properties of light nuclei far from
$\bt$ stability are partly related to the poorly known density
dependence of the symmetry energy.

The charge radius of heavy nuclei such as $^{208}$Pb is little
modified with the change of the density dependence of the symmetry
energy although the neutron radius is apparently
modified\cite{horo1}. Due to the strong couplings of the
neighboring neutron and proton shells in light nuclei, it is not
so simple to say that the charge radius is still  unchanged. In
turn, as the proton skin emerges for the light proton-rich
isotopes, we even conjecture that a relation between the neutron
radius of $^{208}$Pb and the charge radius of the proton-rich
isotope (for instance, $^{12}$O) will be established. It is aim of
this study to constrain the density dependence of the symmetry
energy through the investigation of the charge density
distribution of the light nuclei far from the $\bt$ stability.

The relativistic mean-field (RMF) theory (for reviews,
see{\cite{serot,ring}) is  powerful in describing the ground-state
properties of nuclei from the proton drip line to the neutron drip
line, since it can give dynamically rise to the spin-orbit
interaction associated with shell effects. In this work, we take
into account the isoscalar-isovector terms that modify the density
dependence of the symmetry energy in the framework of the RMF
models. In a RMF model, the $\r NN$ tensor coupling that does not
change any properties of nuclear matter plays its role in finite
nuclei. The interest to include the $\r NN$ tensor coupling is
arisen by the following factors. On one hand, the
isoscalar-isovector terms give modifications to the source term of
the $\r$ meson (i.e., the related density), on the other hand,
the tensor coupling is relevant to the gradient of the related
density. We will investigate oxygen isotopes for their simple
spherical shapes.

The effective Lagrangian density  is given as follows
 \bey
 {\cal L}&=&
{\overline\psi}[i\gm_{\mu}\partial^{\mu}-M_N+g_{\sg}\sg-g_{\om}
\gm_{\mu}\om^{\mu}-g_\r\gm_\mu \tau_3 b_0^\mu\nnb\\
 & &+\frac{f_\r}{2M_N}\sg_{\mu\nu}\pp^\nu b_0^\mu
  -e\frac{1}{2}(1+\tau_3)\gm_\mu A^\mu]\psi-U(\sg,\om^\mu, b_0^\mu)\nnb\\
&& +\frac{1}{2}
(\partial_{\mu}\sg\partial^{\mu}\sg-m_{\sg}^{2}\sg^{2})-
\frac{1}{4}F_{\mu\nu}F^{\mu\nu}+
      \frac{1}{2}m_{\om}^{2}\om_{\mu}\om^{\mu} \nnb\\
      &  &
    - \frac{1}{4}B_{\mu\nu} B^{\mu\nu}+
      \frac{1}{2}m_{\r}^{2} b_{0\mu} b_0^{\mu}-\frac{1}{4}A_{\mu\nu}
      A^{\mu\nu}
\label{eq:lag1}
  \eey
 where $\psi,\sg,\om$, and  $b_0$  are the fields of
the nucleon,  scalar, vector, and charge-neutral isovector-vector
mesons, with their masses $M_N, m_\sg,m_\om$, and $m_\r$,
respectively. $A_\mu$ is the field of the photon.
$g_i(i=\sg,\om,\r)$ and $f_\r$ are the corresponding meson-nucleon
couplings. $\tau$ and $\tau_3$ is the isospin Pauli matrix and its
third component, respectively. $F_{\mu\nu}$,  $ B_{\mu\nu}$, and
$A_{\mu\nu}$   are the strength tensors of the $\om$, $\r$ mesons
and the photon, respectively
\be
F_{\mu\nu}=\pp_\mu \om_\nu -\pp_\nu \om_\mu,\hbox{  }
B_{\mu\nu}=\pp_\mu b_{0\nu} -\pp_\nu b_{0\mu},\hbox{ }
A_{\mu\nu}=\pp_\mu A_{\nu} -\pp_\nu A_{\mu}\ee The
self-interacting terms of $\sigma,\om$-mesons and  the
isoscalar-isovector ones are in the following
 \bey
 U(\sg,\om^\mu, b_0^\mu)&=&\frac{1}{3}g_2\sg^3+\frac{1}{4}g_3\sg^4
 -\frac{1}{4}c_3(\om_\mu\om^\mu)^2,\nnb\\
 &&-4g^2_\r (\Ld_s g_\sg^2\sg^2+\Ld_v g_\om^2
 \om_\mu\om^\mu)b_{0\mu}b_0^\mu
 \eey
Since detailed procedures  are easily found in references (e.g.,
see Ref.\cite{ring}) we only write out the radial Dirac equations
and the tensor potentials from the $\r NN$ tensor coupling
explicitly
 \bey
\frac{dG_\af}{dr}&=&(M^*_N(r)+V(r)+E_\af+U_1^\r(r) )F_\af
-(\frac{\kappa}{r}-U_2^\r(r)-U_3^\r(r))G_\af \label{up}\\
\frac{dF_\af}{dr}&=&(M^*_N(r)-V(r)-E_\af-U_1^\r(r) )G_\af
+(\frac{\kappa}{r}-U_2^\r(r)-U_3^\r(r))F_\af\label{dn}
  \eey
  and
 \bey
U_1^\r(r)&=&-\frac{f_\r}{M_N}g_\r\int_0^\infty dr_1r_1^2
G_\r(r,r_1) \r_T(r_1)\nnb\\
 U_2^\r(r)&=&-\frac{f_\r}{2M_N}g_\r\int_0^\infty
dr_1r_1^2 \frac{dG_\r(r,r_1)}{dr} (\r_3(r_1)- 8g_\r(\Ld_s
g_\sg^2\sg^2+\Ld_v g_\om^2\om^2_0)b_0) \label{eq6}\\
U_3^\r(r)&=&-\frac{f_\r^2}{2M_N^2}\int_0^\infty
dr_1r_1^2\frac{dG_\r(r,r_1)}{dr} \r_T(r_1)\nnb
 \eey
 where
 \be
\r_T(r)=\sum_\af^A\frac{2j_\af+1}{4\pi
r^2}t\frac{d}{dr}(G_\af(r)F_\af(r))
 \ee
 \bey
 M^*_N(r)&=&M_N+V^\sg(r)\nnb\\
 V(r)&=&g_\om(r)\om_0(r)+g_\r b_0(r)t+e(\frac{1+t}{2})A_0\\
 V^\sg(r)&=&g_\sg(r)\sg(r)\nnb
  \eey
 with $\r_3$ the difference of the proton and neutron densities,
t=1 for proton and -1 for neutron, and $G_\r(r,r_1)$ the Green
function of the $\r$ meson. The various density definitions follow
the standard procedure\cite{ring}. The charge density is obtained
by folding the proton density and the proton charge
distribution\cite{horo5} \be \r_c(r)=\int
d^3{\bf{r}}^\prime\frac{\mu^3}{8\pi}\exp
(-\mu|{\bf{r}}-{\bf{r}}^\prime|)\r_p({\bf{r^\prime}}),\hbox{ }
\mu=(0.71)^{1/2}GeV \ee with $\r_p$ the proton density.

The pairing correlation is involved  in non-magic nuclei using
Bardeen-Cooper-Schrief (BCS) theory. We use constant pairing gaps
which are obtained from the prescription of M\"oller and Nix
\cite{mn2}: $\Delta_n=4.8/N^{1/3},\hbox{ } \Delta_p=4.8/Z^{1/3}$
with N and Z the neutron and proton numbers, respectively. For odd
nuclei, the Pauli blocking is considered.

We perform calculations with the NL3 parameter set\cite{nl3} and
distinguish two cases with and without the $\r NN$ tensor coupling
for various isoscalar-isovector coupling $\Ld_v$'s. For
simplicity, the isoscalar-isovector coupling $\Ld_s$ is set as
zero. The symmetry energy at saturation density is not well
constrained, and hence some average of the symmetry energy at
saturation density and the surface energy is constrained by the
binding energy of nuclei. For a given coupling $\Ld_v$, we follow
Ref.\cite{horo1,todd} to readjust the $\r NN$ coupling constant
$g_\r$ so as to keep an average symmetry energy fixed as 25.68 at
$k_F=1.15$ fm$^{-1}$. In doing so, it was found in
Ref.\cite{horo1} that the binding energy of $^{208}$Pb is nearly
unchanged for various $\Ld_v$'s. In our calculation, the total
binding energy of light exotic nuclei with different $\Ld_v$'s is
modified by the same order of $^{208}$Pb. The $\r NN$ tensor
coupling $f_\r$ is taken as $f_\r/g_\r=6.1$ which is used in the
Bonn potentials\cite{broc} and is within $8.0\pm2.0$ obtained from
the light cone sum rules by Zhu\cite{shi}. Note that this ratio of
$f_\rho$ to $g_\rho$, coupled with the relatively large values of
$g_\rho$ in many relativistic mean field models that are usually
fixed by the empirical values of the symmetry energy, could
overestimate the contribution of the rho tensor coupling, and that
will be analyzed below.

 \begin{table}[bh]
\caption{Results for $^{24}$O with the NL3 parameter set for
$f_\r=0$ and $f_\r=6.1g_\r$, respectively. The total binding
energy ($E_b$), the proton radius ($r_p$), and the neutron
thickness ($r_p-r_n$) are given for various $\Ld_v$'s. The
experimental binding energy is 168.48MeV. } \label{tab1}
 \begin{center}
    \begin{tabular}{ c c c c| c c c }
\hline  &\multicolumn{3}{c|}{$f_\r=0$}
 & \multicolumn{3}{c}{$f_\r=6.1g_\r$} \\\cline{2-7}
 $\Ld_v$&$E_b$ (MeV) & $r_p$ (fm) & $r_n-r_p$ (fm) & $E_b$ (MeV) &
  $r_p$ (fm) &$r_n-r_p$ (fm)\\ \hline
  0.000&170.44 & 2.633 & 0.651 &172.64  & 2.639 & 0.619 \\
  0.005&170.14 & 2.635 & 0.644 &172.34  & 2.642 & 0.610 \\
  0.010&169.74 & 2.637 & 0.638 &171.91  & 2.646 & 0.600 \\
  0.015&169.40 & 2.640 & 0.632 &171.42  & 2.650 & 0.591 \\
  0.020&168.93 & 2.643 & 0.626 &170.82  & 2.655 & 0.580 \\
  0.025&168.32 & 2.646 & 0.621 &170.21  & 2.660 & 0.569 \\
  0.030&167.72 & 2.649 & 0.616 &169.43  & 2.667 & 0.557 \\
  \hline
     \end{tabular}
  \end{center}
 \end{table}
At the vicinity of the line $N=Z$, the $\r$ meson together with
the isoscalar-isovector term has almost negligible contribution.
Consequently, we study the isoscalar-isovector term contribution
in nuclei close to the drip lines of the oxygen. Table \ref{tab1}
gives the binding energy, proton radius, neutron thickness for
$^{24}$O with and without the $\r NN$ tensor coupling as $\Ld_v$
is changed. The binding energy of $^{24}$O shows a small reduction
as the $\Ld_v$ is increased, while for $^{208}$Pb the binding
energy shows a weak increase\cite{horo1}. The neutron thickness is
significantly reduced with the increasing $\Ld_v$. Without the
tensor coupling, the proton radius shows a much less modification
compared to the neutron radius, which is similar to that for
$^{208}$Pb\cite{todd}. As the $\r NN$ tensor coupling is included,
an apparent difference appears: the proton radius modification due
to the $\Ld_v$ is largely enhanced, and almost half the
contribution to the decrease of the neutron thickness of
$^{23,24}$O comes from the increase of the proton radius. We
notice that  the inclusion of the tensor coupling in $^{208}$Pb
just brings out negligible changes due to the quite small tensor
potentials.

\begin{figure}[thb]
\begin{center}
\vspace*{-25mm} \epsfig{file=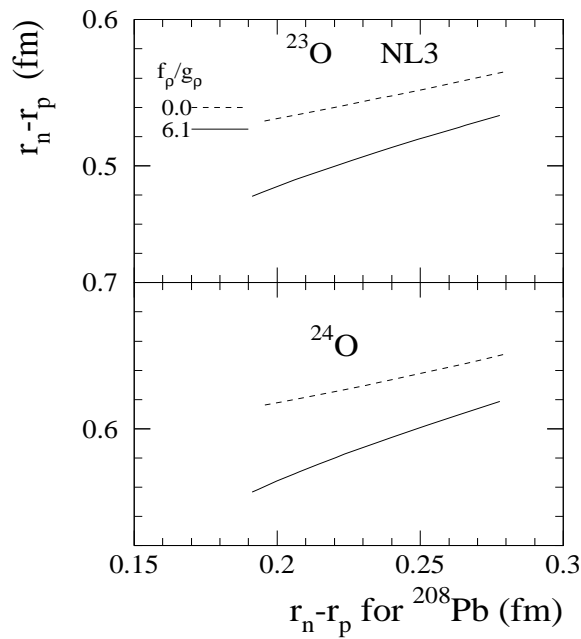,height=12.0cm,width=10.0cm}
 \end{center}
\caption{The correlation of the  neutron thickness of $^{208}$Pb
and that of $^{23,24}$O.\label{fig1}}
\end{figure}

A data  to data correlation between the neutron thickness of
$^{208}$Pb and that of $^{23,24}$O is displayed in Fig.\ref{fig1}.
The correlation is linear, and the $\r NN$ tensor coupling
increases the correlation slope strongly. The relativistic models
with a softer symmetry energy have a larger symmetry energy at low
densities.  It was pointed out in Ref.\cite{todd} that the softer
models are the first ones to drip neutrons that show a diffusive
distribution with the larger valence neutron radius. A similar
result for $^{23,24}$O is obtained that the radius of  the valence
$2s_{1/2}$ neutron increases with softening the symmetry energy.
On the other hand, the change of the neutron radius is mainly from
the modification of core neutron distribution instead of the
diffusive neutron distribution that is often called as the neutron
skins or halos. Consequently, owing to the coupling between the
core neutrons and protons at the neighboring shells in light
isotopes the modification of the neutron density distribution
leads to a similar modification for the proton (charge) density
distribution. In Fig.\ref{fig2}, the charge density for $^{24}$O
is displayed. As seen in Fig.\ref{fig2}, the modification of the
charge density distribution occurs explicitly in the core region
with various $\Ld_v$'s, and  the modification is enhanced due to
the inclusion of the $\r NN$ tensor coupling. Further, we make a
brief analysis of the overestimation of the contribution of the
$\r NN$ tensor coupling due to the relatively large values of
$g_\rho$ in many relativistic mean field models. As we decrease
the value of $f_\r$ by 15\%, the slope of  the neutron thickness
correlation, shown in Fig.\ref{fig1}, is just reduced by 8\%. In
this sense, the importance of the $\r NN$ tensor coupling in
present results is not much overestimated.

\begin{figure}[thb]
\begin{center}
\vspace*{-25mm}
\epsfig{file=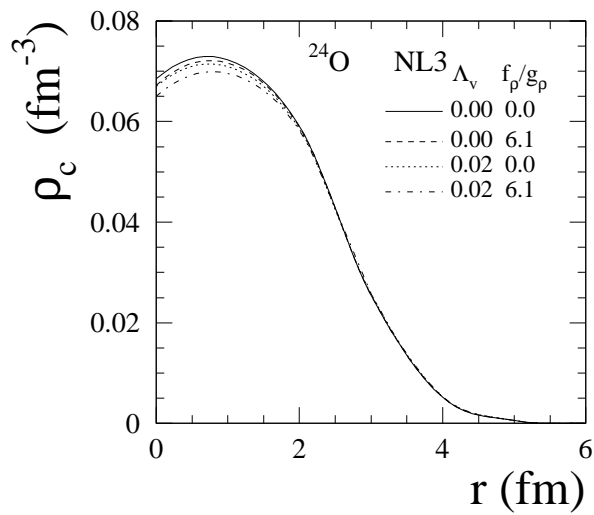,height=10.0cm,width=10.0cm}
 \end{center}
\caption{ The charge density distributions for $^{24}$O with
various cases. \label{fig2} }
\end{figure}

\begin{figure}[thb]
\begin{center}
\vspace*{-25mm} \epsfig{file=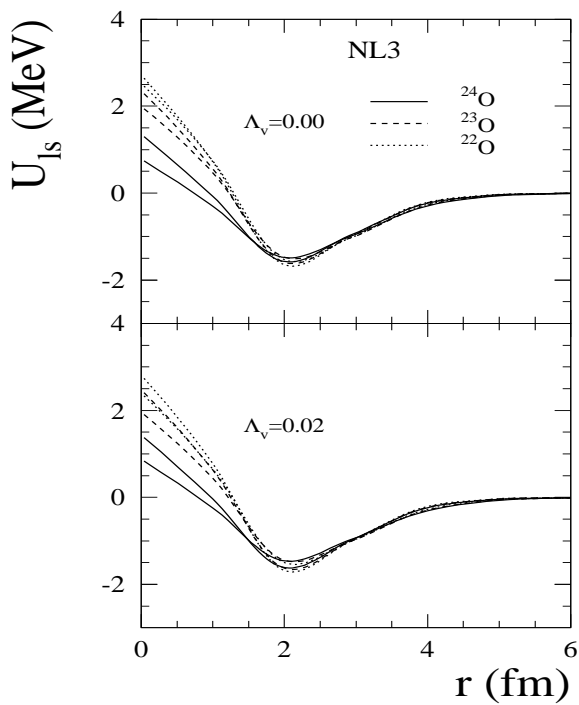,height=12.0cm,width=10.0cm}
\caption{ The neutron spin-orbit potential for $^{22,23,24}$O. The
neutron occupation probability in the orbital $2s_{1/2}$, just
slightly influenced by the $\Ld_v$ and $f_\r$, is 0.35, 0.5, 0.83
for $^{22,23,24}$O, respectively. The lower one of the same type
curves in each panel represents for the case $f_\r/g_\r=6.1$,
while the upper one is for $f_\r/g_\r=0$. \label{fig3} }
\end{center}
\end{figure}

The explicit enhancement of the correlation for $^{23,24}$O as
shown in Fig.\ref{fig1} due to the $\r NN$ tensor coupling does
not occur for isotopes with more neutrons. At the same time, with
the reduction of  the neutron number the contribution from the $\r
NN$ tensor coupling drops. These phenomena can be explained by the
contribution from the orbital $2s_{1/2}$. The orbital $2s_{1/2}$
generates large derivatives of the density distribution with
respect to the radius near the node at the core region. A large
tensor coupling contribution, corresponding to  large tensor
potentials, may come from the large derivatives appearing in
integrals in Eq.(\ref{eq6}).  For isotopes with $N>16$, the
$2s_{1/2}$ neutron distribution is squeezed due to the deeper
binding, and a prominent cancellation occurs in the integrals over
the density derivatives  from the regions inside and outside the
node, leading to dropping tensor potentials. The derivatives
outside the node are small  for $^{24}$O due to the extensive
distribution, and the cancellation  is slight so that large tensor
potentials are produced.  For isotopes with $N<16$, the drop of
the contribution from the $\r NN$ tensor coupling is related to a
smaller occupation probability in $2s_{1/2}$. Fig.\ref{fig3}
displays the neutron spin-orbit potential $U_{ls}$ that is
obtained in an equation for the big component $G_\af$ of the Dirac
spinor by eliminating the small component $F_\af$ from
Eqs.(\ref{up},\ref{dn}). The spin-orbit potential $U_{ls}$ varies
sensitively with the different $2s_{1/2}$ occupation. If the Pauli
blocking is not considered in $^{23}$O, the $2s_{1/2}$ occupation
probability shows a small increment by about 0.1 that may produce
an obvious move  towards the spin-orbit potential for $^{24}$O. As
the occupation probability in $2s_{1/2}$ rises, the difference
arisen by the $\r NN$ tensor coupling goes up, and that
corresponds to the correlation enhancement  observed in
Fig.\ref{fig1}. The difference of the spin-orbit potential exists
mainly in the core region as shown in Fig.\ref{fig3}. This is
consistent with the charge density distributions  shown in
Fig.\ref{fig2}.

\begin{figure}[thb]
\begin{center}
\vspace*{-25mm}
\epsfig{file=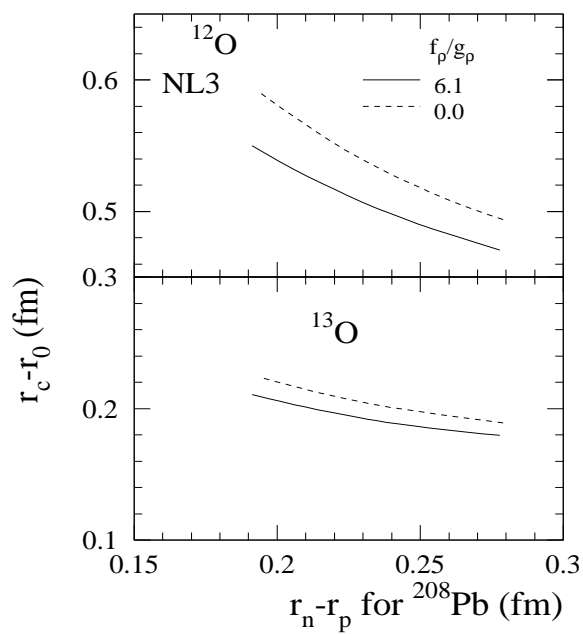,height=12.0cm,width=10.0cm} \caption{ The
correlation of the  neutron thickness of $^{208}$Pb and the charge
radius of $^{12,13}$O.  $r_0$ is the reference radius with
$r_0=3.0$fm. \label{fig4} }
\end{center}
\end{figure}

\begin{figure}[thb]
\begin{center}
\vspace*{-25mm}
\epsfig{file=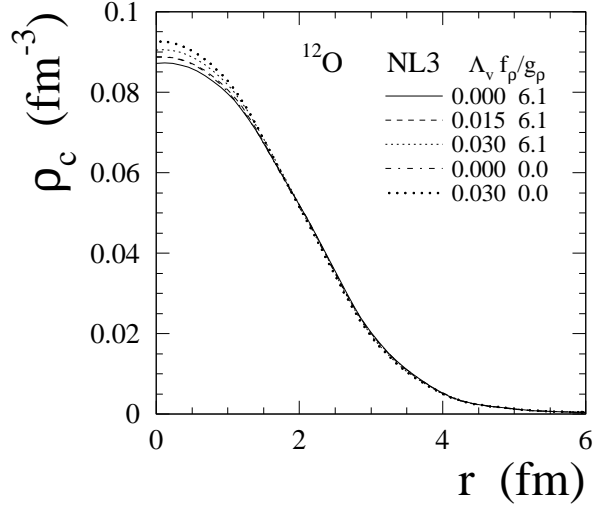,height=10.0cm,width=10.0cm} \caption{ The
charge density distribution for $^{12}$O with various cases.
\label{fig5} }
\end{center}
\end{figure}

Now we investigate the role of the isoscalar-isovector term in the
proton-rich oxygen isotopes. For the binding energy, the
modification due to this term is small. For instance, the binding
energy of $^{12}$O changes by 1.5MeV as the $\Ld_v$ increases from
0 to 0.03. With the decrease of the neutron number, the neutron
distribution is little changed by the isoscalar-isovector term. A
correlation between the charge radius of proton-rich oxygen
isotopes and the neutron thickness of $^{208}$Pb is thus
available. The establishment of the data to data correlation is
helpful since it may bridge the different kind of experiments: the
neutron radius measurement for $^{208}$Pb\cite{jeff} via parity
violating electron scattering and the electron scattering on the
exotic nuclei\cite{riken,gsi}. In Fig.\ref{fig4}, the correlation
between the charge radius of $^{12,13}$O and the neutron thickness
of $^{208}$Pb is shown. The $\r NN$ tensor coupling has a much
less impact on the slope of the correlation relation for
$^{12,13}$O than that for $^{23,24}$O. As different from the
neutron-rich isotopes, the correlation relation for these
proton-rich isotopes shows a nonlinear feature. Contrary to the
correlation relation for the neutron-rich isotopes $^{23,24}$O,
softer models that produce the small neutron thickness for
$^{208}$Pb give rise to larger charge radii for $^{12,13}$O as
shown in Fig.\ref{fig4}.

In Fig.\ref{fig5}, the charge density distributions for $^{12}$O
at different cases  are shown. The dependence on the
isoscalar-isovector term for the charge density distribution is
just slightly affected by the  $\r NN$ tensor coupling. Similar to
that of the neutron-rich isotopes $^{23,24}$O, the change of
charge distribution due to the isoscalar-isovector term is
concentrated in the core region. However,  comparing to
Fig.\ref{fig2},   the rising or dropping trend of the core charge
density distribution is different with increasing $\Ld_v$. As
shown in Fig.\ref{fig4}, the charge density distribution of the
drip line isotope ($^{12}$O) is more sensitive to the change of
the density dependence of the symmetry energy. This may be
attributed to the explicit increasing mean-field expectation value
of the isovector meson for isotopes towards the proton drip line.
The charge density distribution can be related to the cross
section of the electron scattering on these exotic nuclei. As
pointed out in Ref.\cite{wang,sick}, the form factor in the
eikonal approximation at high momentum transfer (about
3.0fm$^{-1}$) is sensitive to the charge distribution at the core
region. According to the calculated charge density distributions
in some oxygen isotops, the identification of the
isoscalar-isovector terms requires  the electron scattering on
exotic nuclei with  high momentum transfers.

We have not made some extensive discussions on the model
dependence of the results based on various RMF parameter sets
instead of using the NL3 parameter set, since  a detailed
discussion on the model dependence may be found elsewhere(for
instances, see \cite{horo1,todd}). For the RMF model the variation
of the neutron radius of $^{208}$Pb is accomplished by softening
the symmetry energy through introducing the isoscalar-isovector
term that does not change a variety of ground-state properties.
The different theoretical models may extend the region of the
uncertainty for the neutron radius of $^{208}$Pb \cite{horo1,todd}
that is  about 0.3fm. From the correlation between the charge
radius of some exotic oxygen isotopes and the neutron thickness of
$^{208}$Pb, we find there are also considerable uncertainties in
the charge radius of exotic oxygen isotopes that awaits
experimental identification. For instance, even according to an
estimation based on a linear extrapolation, the uncertainty of the
$^{12}$O charge radius that is almost from the core region is
about 0.3fm.

In summary, we have studied the charge density distributions of
some exotic oxygen isotopes in RMF. The isoscalar-isovector term
which is not constrained by the present data is considered to
simulate the uncertainty of the neutron thickness of $^{208}$Pb.
The $\r NN$ tensor coupling that does not affect the symmetry
energy in the mean field approximation is included. Strong
correlations between the charge radius of $^{12,13,23,24}$O and
the neutron thickness of $^{208}$Pb are established,  showing that
charge radii of $^{12,13,23,24}$O, especially $^{12}$O, exist a
considerable uncertainty.  The neutron thickness for $^{23,24}$O
reduces with softening the symmetry energy.  The strong
correlation between the radii of $^{23,24}$O and the neutron
radius of $^{208}$Pb can be largely attributed to their special
structure and the tensor coupling: The $\r NN$ tensor coupling
explicitly enhances the correlation due to the neutron occupation
of the  orbital $2s_{1/2}$. The modifications of the charge
density distributions for these exotic isotopes due to the
isoscalar-isovector term are mainly explicit at the core region.
The charge radius is sensitive to the change of the density
dependence of the symmetry energy as the isotope goes towards the
proton drip line. These new correlations may bridge two kind of
experiments, the neutron radius measurement for $^{208}$Pb and the
charge distribution measurement for exotic nuclei, to jointly
constrain the density dependence of the symmetry energy.

\section*{Acknowledgement}
This work is partially supported by CAS Knowledge Innovation
Project No.KJCX2-N11, the National Natural Sciences Foundation of
China under grant No.10405031, 10235030 and the Major State Basic
Research Development Program under grant No. G200077400.

\end{document}